\documentclass[aps,twocolumn,prd,nofootinbib,superscriptaddress]{revtex4}
\usepackage{amsmath}
\usepackage{graphicx}
\usepackage{dcolumn}
\usepackage{bm}
\usepackage{amssymb}
\usepackage{latexsym}

\newcommand{\be}{\begin{equation}}
\newcommand{\ee}{\end{equation}}
\newcommand{\bq}{\begin{eqnarray}}
\newcommand{\eq}{\end{eqnarray}}

\begin{document}

%\preprint{McGill 04-xxx}

\title{Holographic tachyon model}

\author{Jingfei Zhang}
\affiliation{School of Physics and Optoelectronic Technology, Dalian
University of Technology, Dalian 116024, People's Republic of China}
\author{Xin Zhang}
\affiliation{Kavli Institute for Theoretical Physics China,
Institute of Theoretical Physics, Chinese Academy of Sciences
(KITPC/ITP-CAS), P.O.Box 2735, Beijing 100080, People's Republic of
China}
\author{Hongya Liu}
\affiliation{School of Physics and Optoelectronic Technology, Dalian
University of Technology, Dalian 116024, People's Republic of China}

\begin{abstract}
We propose in this Letter a holographic model of tachyon dark
energy. A connection between the tachyon scalar-field and the
holographic dark energy is established, and accordingly, the
potential of the holographic tachyon field is constructed. We show
that the holographic evolution of the universe with $c\geqslant 1$
can be described completely by the resulting tachyon model in a
certain way.

\end{abstract}

\maketitle

The astronomical observations over the past decade indicate that our
universe is currently undergoing an epoch of accelerated expansion
(see e.g. Refs. \cite{SN,CMB,LSS}). Such an accelerated expansion
implies, following the Friedmann-Robertson-Walker (FRW) cosmology,
the existence of a ``dark energy'', a mysterious exotic matter with
large enough negative pressure, whose energy density has been a
dominative power of the universe. The astrophysical feature of dark
energy is that it remains unclustered at all scales where
gravitational clustering of baryons and nonbaryonic cold dark matter
can be seen.%\footnote{This statement is of course model-dependent:
%it is true in quintessence scenarios for example, but not true in
%chameleon models or mass-varying neutrino models, for example.}
Its gravity effect is shown as a repulsive force so as to make the
expansion of the universe accelerate when its energy density becomes
dominative power of the universe. The combined analysis of
cosmological observations suggests that the universe is spatially
flat, and consists of about $70\%$ dark energy, $30\%$ dust matter
(cold dark matter plus baryons), and negligible radiation. The most
obvious candidate for the dark energy is the cosmological constant
$\lambda$ \cite{Einstein:1917} for which $w=-1$ (for reviews see
e.g. Refs. \cite{cc}). However, the cosmological constant always
suffers from the ``fine-tuning'' and ``cosmic coincidence''
problems. Another candidate for dark energy is the energy density
associated with dynamical scalar-field, a slowly varying, spatially
homogeneous component. An example of scalar-field dark energy is the
so-called ``quintessence'' \cite{quintessence}, a scalar field $Q$
slowly evolving down its potential $V(Q)$. Provided that the
evolution of the field is slow enough, the kinetic energy density is
less than the potential energy density, giving rise to the negative
pressure responsible to the cosmic acceleration. So far a wide
variety of scalar-field dark energy models have been proposed.
Besides quintessence, these also include phantom \cite{phantom},
$K$-essence \cite{kessence}, tachyon \cite{tachyon}, ghost
condensate \cite{ghost} and quintom \cite{quintom} amongst many. But
we should note that the mainstream viewpoint regards the scalar
field dark energy models as an effective description of an
underlying theory of dark energy. In addition, other proposals on
dark energy include interacting dark energy models \cite{intde},
braneworld models \cite{brane}, and Chaplygin gas models \cite{cg},
etc.. One should realize, nevertheless, that almost these models are
settled at the phenomenological level, lacking theoretical root.

In recent years, many string theorists have devoted to understand
and shed light on the cosmological constant or dark energy within
the string framework. The famous Kachru-Kallosh-Linde-Trivedi (KKLT)
model \cite{kklt} is a typical example, which tries to construct
metastable de Sitter vacua in the light of type IIB string theory.
Furthermore, string landscape idea \cite{landscape} has been
proposed for shedding light on the cosmological constant problem
based upon the anthropic principle and multiverse speculation.
Another way of endeavoring to probe the nature of dark energy within
the fundamental theory framework originates from some considerations
of the features of the quantum gravity theory. It is generally
believed by theorists that we can not entirely understand the nature
of dark energy before a complete theory of quantum gravity is
established. However, although we are lacking a quantum gravity
theory today, we still can make some attempts to probe the nature of
dark energy according to some principles of quantum gravity. The
holographic dark energy model
\cite{Cohen:1998zx,Hsu:2004ri,Li:2004rb} is just an appropriate
example, which is constructed in the light of the holographic
principle \cite{holoprin} of quantum gravity theory. That is to say,
the holographic dark energy model possesses some significant
features of an underlying theory of dark energy.

According to the holographic principle, the number of degrees of
freedom for a system within a finite region should be finite and
should be bounded roughly by the area of its boundary. In the
cosmological context, the holographic principle will set an upper
bound on the entropy of the universe. Motivated by the Bekenstein
entropy bound, it seems plausible to require that for an effective
quantum field theory in a box of size $L$ with UV cutoff $\Lambda$,
the total entropy should satisfy $S=L^3\Lambda^3\leqslant
S_{BH}\equiv\pi M_P^2L^2$, where $S_{BH}$ is the entropy of a black
hole with the same size $L$. However, Cohen et al.
\cite{Cohen:1998zx} pointed out that to saturate this inequality
some states with Schwartzschild radius much larger than the box size
have to be counted in. As a result, a more restrictive bound, the
energy bound, has been proposed to constrain the degrees of freedom
of the system, requiring that the total energy of a system with size
$L$ should not exceed the mass of a black hole with the same size,
namely, $L^3\Lambda^4=L^3\rho_\Lambda\leqslant L M_P^2$. This means
that the maximum entropy is in order of $S_{BH}^{3/4}$. When we take
the whole universe into account, the vacuum energy related to this
holographic principle is viewed as dark energy, usually dubbed
holographic dark energy. The largest IR cut-off $L$ is chosen by
saturating the inequality so that we get the holographic dark energy
density
\begin{equation}
\rho_{\Lambda}=3c^2M_P^2L^{-2}~,\label{de}
\end{equation} where $c$ is a numerical constant, and $M_P\equiv 1/\sqrt{8\pi
G}$ is the reduced Planck mass. Many authors have devoted to develop
the idea of the holographic dark energy (see e.g. Refs.
\cite{holoext}). It has been conjectured by Li \cite{Li:2004rb} that
the IR cutoff $L$ should be given by the future event horizon of the
universe
\begin{equation}
R_{\rm eh}(a)=a\int\limits_t^\infty{dt'\over
a(t')}=a\int\limits_a^\infty{da'\over Ha'^2}~.\label{eh}
\end{equation}
Such a holographic dark energy looks reasonable, since it may
provide simultaneously natural solutions to both dark energy
problems as demonstrated in Ref.\cite{Li:2004rb}. The holographic
dark energy model has been tested and constrained by various
astronomical observations \cite{obs}.

The holographic dark energy scenario reveals the dynamical nature of
the vacuum energy. When taking the holographic principle into
account, the vacuum energy density will evolve dynamically. On the
other hand, let us consider another dynamical dark energy candidate,
the scalar-field dark energy. The scalar field dark energy models
are often viewed as effective description of the underlying theory
of dark energy. However, the underlying theory of dark energy can
not be achieved before a complete theory of quantum gravity is
established. We can, nevertheless, speculate on the underlying
theory of dark energy by taking some principles of quantum gravity
into account. The holographic dark energy model is no doubt a
tentative in this way. We are now interested in that if we assume
the holographic vacuum energy scenario as the underlying theory of
dark energy, how the scalar field model can be used to effectively
describe it. In this direction, some work has been done. The issues
of holographic quintessence and holographic quintom have been
discussed in Refs. \cite{holoquin} and \cite{hologhost}. In this
Letter, we will construct the holographic tachyon model, connecting
the tachyon scalar-field with the holographic dark energy.

The rolling tachyon condensate in a class of string theories may
have interesting cosmological consequences. It has been shown by Sen
\cite{tachyon} that the decay of D-branes produces a pressureless
gas with finite energy density that resembles classical dust. A
rolling tachyon has an interesting equation of state whose parameter
smoothly interpolates between $-1$ and $0$ \cite{tachyon2}. Thus,
tachyon can be viewed as a suitable candidate for the inflaton at
high energy \cite{tachinfl}. Meanwhile, the tachyon can also act as
a source of dark energy depending upon the form of the tachyon
potential \cite{tachde}. We shall consider a tachyon model with
definite holography nature in this Letter. In what follows we shall
construct the holographic tachyon potential according to the
holographic evolution of the universe. The effective Lagrangian for
the tachyon on a non-BPS D3-brane is described by
\begin{equation}
S=-\int d^4xV(\phi)\sqrt{-{\rm
det}(g_{ab}+\partial_a\phi\partial_b\phi)},
\end{equation}
where $V(\phi)$ is the tachyon potential. The corresponding energy
momentum tensor has the form
\begin{equation}
T_{\mu\nu}={V(\phi)\partial_\mu\phi\partial_\nu\phi\over
\sqrt{1+g^{\alpha\beta}\partial_\alpha\phi\partial_\beta\phi}}
-g_{\mu\nu}V(\phi)\sqrt{1+g^{\alpha\beta}\partial_\alpha\phi\partial_\beta\phi}.
\end{equation}
In a flat FRW background the energy density $\rho_{\rm t}$ and the
pressure density $p_{\rm t}$ are given by
\begin{equation}
\rho_{\rm t}=-T^0_0={V(\phi)\over
\sqrt{1-\dot{\phi}^2}},\label{rhot}
\end{equation}
\begin{equation}
p_{\rm t}=T^i_i=-V(\phi)\sqrt{1-\dot{\phi}^2}.\label{pt}
\end{equation}
The equation of state of the tachyon is consequently given by
\begin{equation}
w_{\rm t}=p_{\rm t}/\rho_{\rm t}=\dot{\phi}^2-1.\label{wt}
\end{equation}
We see that irrespective the steepness of the tachyon potential, the
equation of state varies between 0 and $-1$. Clearly, the tachyonic
scalar field cannot realize the equation of state crossing $-1$.

Imposing the holographic nature to the tachyon, the energy density
of tachyon is needed to satisfy the requirement of holographic
principle, i.e., we should identify $\rho_{\rm t}$ with
$\rho_\Lambda$. It is remarkable that here the condition $c\geqslant
1$ is needed as will be discussed below. Consider a universe filled
with matter component $\rho_{\rm m}$ (including both baryon matter
and cold dark matter) and holographic tachyon component $\rho_{\rm
t}$, the Friedmann equation reads
\begin{equation}
3M_{P}^2H^2=\rho_{\rm m}+\rho_{\rm t},
\end{equation} or equivalently,
\begin{equation}
H(z)=H_0\left(\Omega_{\rm m0}(1+z)^3\over 1-\Omega_{\rm
t}\right)^{1/2},\label{Ez}
\end{equation} where $z=(1/a)-1$ is the redshift of the universe.
Note that we always assume spatial flatness throughout this Letter
as motivated by inflation. Combining the definition of the
holographic dark energy (\ref{de}) and the definition of the future
event horizon (\ref{eh}), we derive
\begin{equation}
\int\limits_a^\infty{d\ln a'\over Ha'}={c\over Ha\sqrt{\Omega_{\rm
t}}}.\label{rh}
\end{equation} We notice that the Friedmann
equation (\ref{Ez}) implies
\begin{equation}
{1\over Ha}=\sqrt{a(1-\Omega_{\rm t})}{1\over H_0\sqrt{\Omega_{\rm
m0}}}.\label{fri}
\end{equation} Substituting (\ref{fri}) into (\ref{rh}), one
obtains the following equation
\begin{equation}
\int\limits_x^\infty e^{x'/2}\sqrt{1-\Omega_{\rm t}}dx'=c
e^{x/2}\sqrt{{1\over\Omega_{\rm t}}-1}~,
\end{equation} where $x=\ln a$. Then taking derivative with respect to $x$ in both
sides of the above relation, we get easily the dynamics satisfied by
the dark energy, i.e. the differential equation about the fractional
density of dark energy,
\begin{equation}
\Omega'_{\rm t}=-(1+z)^{-1}\Omega_{\rm t}(1-\Omega_{\rm
t})\left(1+{2\over c}\sqrt{\Omega_{\rm t}}\right),\label{deq}
\end{equation}
where the prime denotes the derivative with respect to the redshift
$z$. This equation describes behavior of the holographic dark energy
completely, and it can be solved exactly \cite{Li:2004rb}. From the
energy conservation equation of the dark energy, the equation of
state of the dark energy can be given
\begin{equation}
w_{\rm t}=-1-{1\over 3}{d\ln\Omega_{\rm t}\over d\ln a}=-{1\over
3}(1+{2\over c}\sqrt{\Omega_{\rm t}})~.\label{w}
\end{equation} Note that the formula
$\rho_{\rm t}={\Omega_{\rm t}\over 1-\Omega_{\rm t}}\rho_{\rm
m}^0a^{-3}$ and the differential equation of $\Omega_{\rm t}$
(\ref{deq}) are used in the second equal sign. Using Eqs.
(\ref{rhot}), (\ref{wt}) and (\ref{Ez}), we derive the holographic
tachyon potential
\begin{equation}
{V(\phi)\over \rho_{c0}}={\Omega_{\rm t}\Omega_{\rm m0} (1+z)^3\over
1-\Omega_{\rm t}}\sqrt{-w_{\rm t}},\label{potential}
\end{equation}
where $\Omega_{\rm t}$ and $w_{\rm t}$ are given by Eqs. (\ref{deq})
and (\ref{w}), $\rho_{\rm c0}=3M_P^2H_0^2$ is today's critical
density of the universe. Furthermore, using Eqs. (\ref{wt}) and
(\ref{Ez}), the derivative of the holographic tachyon scalar-field
$\phi$ with respect to the redshift $z$ can be given
\begin{equation}
{\phi'\over H_0^{-1}}=\pm\sqrt{{(1-\Omega_{\rm t})(1+w_{\rm t})\over
\Omega_{\rm m0}(1+z)^5}},\label{phiprime}
\end{equation}
where the sign is actually arbitrary since it can be changed by a
redefinition of the field, $\phi\rightarrow -\phi$. Consequently, we
can easily obtain the evolutionary form of the holographic tachyon
field
\begin{equation}
\phi(z)=\int\limits_0^z\phi'dz,\label{phi}
\end{equation}
by fixing the field amplitude at the present epoch ($z=0$) to be
zero, $\phi(0)=0$.

%%%%%%%%%%%%%%%%%%%%%%%%%%%%%%%%%%%%%%%%%%%%%%%%%%%%%%%%%%%%%%%%%%
\begin{figure}[htbp]
\begin{center}
\includegraphics[scale=0.90]{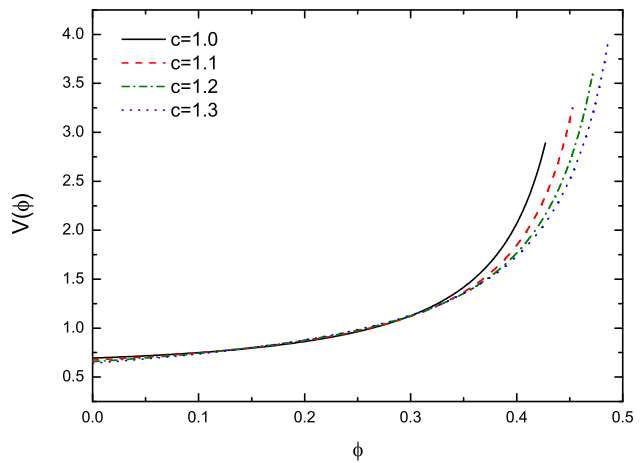}
\caption[]{\small The potential for the holographic tachyon, where
$\phi$ is in unit of $H_0^{-1}$ and $V(\phi)$ in $\rho_{\rm co}$. We
take here $\Omega_{\rm m0}=0.27$.}\label{fig:potential}
\end{center}
\end{figure}
%%%%%%%%%%%%%%%%%%%%%%%%%%%%%%%%%%%%%%%%%%%%%%%%%%%%%%%%%%%%%%%%%%%

%%%%%%%%%%%%%%%%%%%%%%%%%%%%%%%%%%%%%%%%%%%%%%%%%%%%%%%%%%%%%%%%%%
\begin{figure}[htbp]
\begin{center}
\includegraphics[scale=0.90]{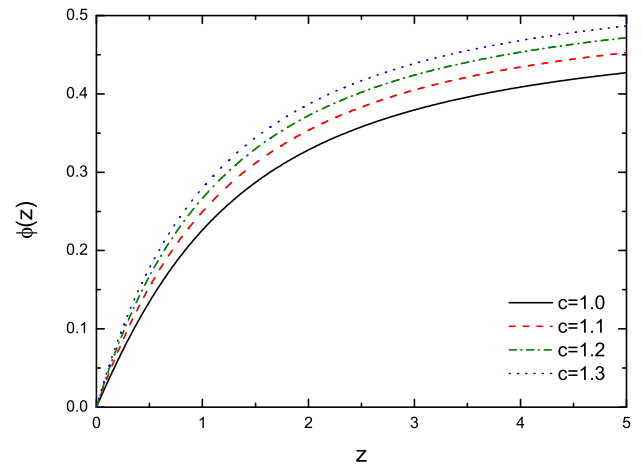}
\caption[]{\small The evolution of the scalar-field $\phi(z)$ for
the holographic tachyon, where $\phi$ is in unit of $H_0^{-1}$. We
take here $\Omega_{\rm m0}=0.27$.}\label{fig:field}
\end{center}
\end{figure}
%%%%%%%%%%%%%%%%%%%%%%%%%%%%%%%%%%%%%%%%%%%%%%%%%%%%%%%%%%%%%%%%%%%

The property of the holographic dark energy is mainly governed by
the numerical parameter $c$. From Eq. (\ref{w}), it can be easily
found that the evolution of the equation of state satisfies
$-(1+2/c)/3\leqslant w\leqslant -1/3$ due to $0\leqslant\Omega_{\rm
t}\leqslant 1$. Thus, the parameter $c$ plays a significant role in
the holographic evolution of the universe. When $c<1$, the
holographic evolution will make the equation of state cross $w=-1$
(from $w>-1$ evolves to $w<-1$); when $c\geqslant 1$, the equation
of state will evolve in the region of $-1\leqslant w\leqslant -1/3$.
Since the equation of state of tachyon scalar-field evolves within
the range of $-1<w<0$, only the holographic evolution of cases
$c\geqslant 1$ can be described by the tachyon. So, it is notable
that the constructed holographic tachyon, Eqs.
(\ref{potential})-(\ref{phi}), must satisfy the condition
$c\geqslant 1$. In fact, in the holographic scenario, the value of
$c$ should be determined by cosmological observations. However,
current observational data cannot determine the value of $c$
accurately due to the precision of these data. An analysis of the
latest observational data, including the gold sample of 182 SNIa,
the CMB shift parameter given by the 3-year WMAP observations, and
the BAO measurement from the SDSS, shows that the possibilities of
$c>1$ and $c<1$ both exist and their likelihoods are almost equal
within 3 sigma error range \cite{holonewobs}.

The tachyon models with different potential forms have been
discussed widely in the literature. For the holographic tachyon
model constructed in this Letter, the potential $V(\phi)$ can be
determined by Eqs. (\ref{potential})-(\ref{phi}). The analytical
form of the potential $V(\phi)$ cannot be derived due to the
complexity of these equations, but we can obtain the holographic
tachyon potential numerically. Using the numerical method, the
holographic tachyon potential $V(\phi)$ is plotted in figure
\ref{fig:potential}, where $\phi(z)$ is also obtained according to
Eqs. (\ref{phiprime}) and (\ref{phi}), also displayed in figure
\ref{fig:field}. Selected curves are plotted for the cases of
$c=1.0$, 1.1, 1.2 and 1.3, and the present fractional matter density
is chosen to be $\Omega_{\rm m0}=0.27$. From figures
\ref{fig:potential} and \ref{fig:field}, we can see the dynamics of
the tachyon scalar field explicitly. According to the holographic
evolution of the universe, the tachyon potential is more steep in
the early epoch $(z\sim 5)$ and becomes very flat near today.
Consequently, the tachyon scalar field $\phi$ rolls down the
potential with the kinetic energy $\dot{\phi}^2$ gradually
decreasing. The equation of state of the tachyon $w_{\rm t}$,
accordingly, decreases gradually with the cosmic evolution, and as a
result $dw_{\rm t}/d\ln a<0$. This feature is very similar to the
holographic quintessence, see Ref.\cite{holoquin} for details.

In summary, we have proposed in this Letter a holographic model of
tachyon dark energy. We adopt the viewpoint of that the scalar field
models of dark energy are effective theories of an underlying theory
of dark energy. The underlying theory, though has not been achieved
presently, is presumed to possess some features of a quantum gravity
theory, which can be explored speculatively by taking into account
the holographic principle of quantum gravity theory. If we regard
the tachyon scalar-field as an effective description of the
underlying theory of dark energy, it should, presumably, carry some
holographic feature. Naturally, the tachyon with holographic feature
should be capable of realizing the holographic evolution of the
universe. We show that the holographic evolution of the universe
with $c\geqslant 1$ can be described completely by the tachyon in a
certain way. A connection between the tachyon and the holographic
dark energy has been established, and the potential of the
holographic tachyon has been constructed accordingly.

\section*{Acknowledgements}

This work is supported by the China Postdoctoral Science Foundation,
the K. C. Wong Education Foundation (Hong Kong), the National
Natural Science Foundation of China, as well as the National Basic
Research Program of China (2003CB716300).

\vskip 0.2cm

{\it Note Added:} During the submission and review process of this
manuscript, Ref. \cite{Setare:2007hq} appeared on the arXiv which
discusses the similar topic to our study.

\end{document}